\documentclass[sigplan,screen]{acmart}

\AtBeginDocument{%
  }

\setcopyright{acmcopyright}
\copyrightyear{2018}
\acmYear{2018}
\acmDOI{XXXXXXX.XXXXXXX}

\acmConference[ICBRA 2023]{10th International Conference on Bioinformatics Research and Applications}{Sep 22-24, 2023}{Barcelona, Spain}
\acmPrice{15.00}
\acmISBN{978-1-4503-XXXX-X/18/06}

\begin{document}

\title{Uncertainty Quantification for Eosinophil Segmentation}

\author{Kevin Lin}
\orcid{0000-0002-4968-0532}
\affiliation{%
  \department{School of Data Science}
  \institution{University of Virginia}
  \city{Charlottesville}
  \state{Virginia}
  \country{USA}
}
\email{pex7ps@Virginia.edu}

\author{Donald Brown}
\orcid{0000-0002-9140-2632}
\affiliation{%
  \department{School of Data Science}
  \institution{University of Virginia}
  \city{Charlottesville}
  \state{Virginia}
  \country{USA}}
\email{deb@virginia.edu}

\author{Sana Syed}
\orcid{0000-0003-0954-0583}
\affiliation{%
  \department{Department of Pediatric Gastroenterology}
  \institution{University of Virginia}
  \city{Charlottesville}
  \state{Virginia}
  \country{USA}
}
\email{ss8xj@virginia.edu}

\author{Adam Greene}
\orcid{0009-0004-3979-5813}
\affiliation{%
  \department{Department of Pediatric Gastroenterology}
  \institution{University of Virginia}
  \city{Charlottesville}
  \state{Virginia}
  \country{USA}
}
\email{arg7ef@virginia.edu}

\renewcommand{\shortauthors}{Lin et al.}

\begin{abstract}
    Eosinophilic Esophagitis (EoE) is an allergic condition increasing in prevalence. To diagnose EoE, pathologists must find 15 or more eosinophils within a single high-power field (400X magnification). Determining whether or not a patient has EoE can be an arduous process and any medical imaging approaches used to assist diagnosis must consider both efficiency and precision. We propose an improvement of Adorno et al's approach for quantifying eosinphils using deep image segmentation. Our new approach leverages Monte Carlo Dropout, a common approach in deep learning to reduce overfitting, to provide uncertainty quantification on current deep learning models. The uncertainty can be visualized in an output image to evaluate model performance, provide insight to how deep learning algorithms function, and assist pathologists in identifying eosinophils. 
\end{abstract}


\received{29 Jun 2023}
\received[revised]{XX XXXX 20XX}
\received[accepted]{XX XXXX 20XX}

\maketitle

\section{Introduction}
\begin{figure}
        \centering
        \includegraphics[width = 50mm]{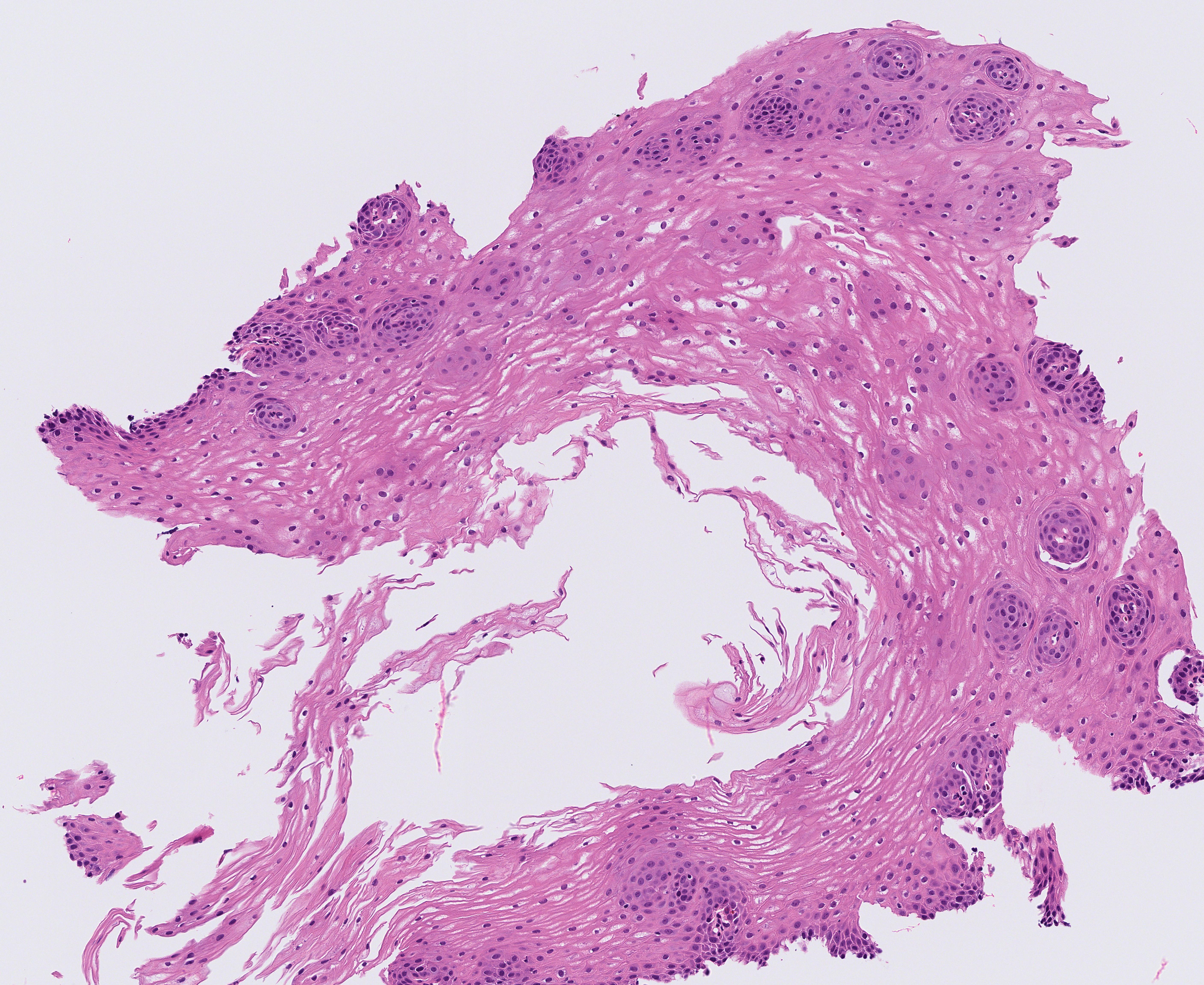}
        \caption{Example Image Data from Gastroenterology Data Science Lab}
        \label{fig:EoE}
    \end{figure}

    Eosinophilic esophagitis (EoE) is an inflammatory disease of the esophagus characterized by the prevalence of a type of white blood cell (eosinophil). Approximately 0.5-1.0 in 1,000 people have EoE and it can be seen in 2-7\% of patients that undergo endoscopies \cite{Dellon2014EpidemiologyEsophagitis}. Although the cause of EoE remains unclear, pathologists believe EoE to be triggered by a patient's diet. Furthermore, EoE is only increasing in prevalence \cite{Carr2018EosinophilicEsophagitis} leading to an increased load on pathologists. Patients with EoE typically present with swallowing difficulties, food impaction, and chest pain \cite{Runge2017CausesEsophagitis}. To diagnose EoE, patients must undergo an endoscopy where eosinophils biopsy tissue samples are then evaluated for concentration of eosinophils. Pathologists diagnose the patient with EoE if at least one High-Power Field (HPF; 400× magnification adjustment) within a patient’s tissue biopsy slide contains 15 or more eosinophils \cite{Furuta2007EosinophilicGastroenterol}. The dataset is obtained from the Gastroenterology Data Science Lab from UVA Hospital patient data. A sample image is given in Figure \ref{fig:EoE}. Each image is 512x512x3 large and there are 514 images/masks in the dataset spanning 30 UVA Medical Center patients. We will keep the three channels [r,g,b] for the image but will import the masks as grayscale. All data is obtained from subjects under conditions of academic use only. No personal health information (PHI) is present in the data. 

\subsection{Related Work}
 The frequency of use of ResNet and UNet models to detect diseases has been increasing in recent years; classification and segmentation of medical images using these models has shown efficient results \cite{Hann2021EnsembleDatasets}\cite{Sarwinda2021DeepCancer}. Namely, Adorno, et al.\cite{Adorno2021AdvancingVision} demonstrated impressive performance in segmentation of Eosinophils to assist diagnosis of Eosinophilic Esophagitis. However, applications of these models in deep neural networks have caused concern among scholars about their inability to measure uncertainty \cite{Begoli2019TheMaking}. There are two common approaches to quantifying uncertainty in machine learning: aleatoric and epistemic. Aleatoric uncertainty refers to the variability in outcomes with random effects, and epistemic uncertainty refers to a lack of information in applying models \cite{Hullermeier2021AleatoricMethods}. For example, while one could include more weights in the neural network to address epistemic uncertainty, aleatoric uncertainty cannot be addressed, only identified. Here we attempted to quantify aleatoric uncertainty using MCD UNet, which has proven effective in other applications without changing or optimizing architectures \cite{Avci2021QuantifyingMRI}. Motivation for this work stems from the drive in medical image analysis to increase efficiency and precision in results. Increasing the efficiency of a model and thereby reducing the time needed to train disease detection models can reduce the load on medical providers and lead to better overall patient outcomes while increasing the precision in the results can reduce the probability of error. In the case where a model is attempting to detect a disease and potentially affect a patient's treatment, there is little room for error.

\section{Methodology}

\subsection{Mathematical Linkage Between Problem and Method}
We will treat the data as discrete since the image is made up of a three dimensional tuple $[r,g,b]$. Since we have a very large dataset, we will use variational approximation, specifically minimizing KL divergence, in order to approximate the distribution of the affected cells. KL divergence is given by
\begin{align*}
    \textbf{KL}&(q(\textbf{Z}|| p(\textbf{Z}|D))) = \\
    -&(E_q(\log p(D,\textbf{Z}))- E_q(\log q(\textbf{Z}))) + \log p(D)
\end{align*}

Since maximizing the evidence lower bound (ELBO) is largely impractical, we will be using a Monte-Carlo approximation. We will use a UNet convolutional neural network architecture and use Monte Carlo (MC) Dropout as a Bayesian approximation to identify which image segments correspond to different cells \cite{Gal2016DropoutLearning}. In Bayesian neural networks, each weight is represented by a probability distribution which we will assume is Gaussian instead of just a number. The learning aspect corresponds to Bayesian inference which we will use MC Sampling. We will measure uncertainty for every pixel using cross-entropy over two classes of "background" $(C=0)$ and "foreground"$(C=1)$: 
\begin{equation*}
    U = -(p_{C=0}\cdot\ln(p_{C=0})+p_{C=1}\cdot \ln(p_{C=1}))
\end{equation*}

We will also explore if we can use aleatoric uncertainty to measure performance. This is given by 
\begin{align*}
    E_{p(z|D)}&H[p(y|z,x)] = \\
    - &\int p(z|D)\left(\sum_{y\in Y} p(y|z,x) \log p(y|z,x)dw\right)
\end{align*}
\subsection{Bayesian Methods}
Work from Gal and Ghahramani in 2016 \cite{Gal2016DropoutLearning}, suggest that a Monte Carlo Dropout UNet is equivalent to the deep Gaussian process used in Bayesian Neural Networks. Essentially, we can minimize the KL divergence using approximation through Monte Carlo integration to get an unbiased estimate. Minimizing the KL divergence between the approximate posterior $q(w)$ and the posterior of the full deep Gaussian Process $p(w|X,Y)$ is given by the objective function:
\begin{equation*}
    -\int q(w)\log p(Y|X,w)dw + KL(q(w)||p(w))
\end{equation*}
The first and second term can be represented by a sum and approximated by Monte Carlo integration. For the Monte Carlo Dropout UNet, we apply dropout before every weight where dropout is defined as switching off neurons at each training step.     

\section{Results}

    The UNet is convolutional network architecture for fast and precise segmentation of images \cite{Ronneberger2015U-Net:Segmentation}. It has been shown to outperform what was previously considered the best method (a sliding-window convolutional network) on the ISBI challenge for segmentation of neuronal structures in electron microscopic stacks. For the UNet model, we used 23 convolutional layers with batch normalization. In both the encoder and decoder steps, we used a ReLU activation function, and for the final layer we used a sigmoid activation function. Loss was computed using binary cross entropy and ADAM was used as the optimizer. To prevent overfitting, we implemented early stopping and data augmentation. The four models we used are MCD UNet, UNet, DenseNet, and ResNet50. All training was done on 4 NVIDIA A100 GPUs with 300GB of RAM in TensorFlow/Keras 2.7. Each model run was ran for 100 epochs with a learning rate of 0.001.

\begin{figure}
        \centering
        \includegraphics[width = 80mm]{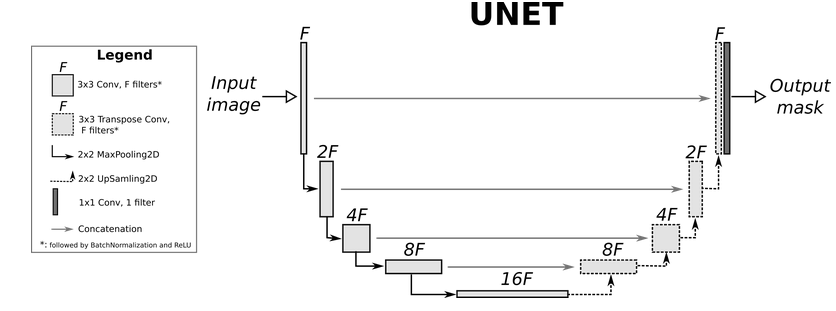}
        \caption{UNet Architecture}
        \label{fig:UNet}
    \end{figure}

    The underlying architecture of the UNet model is shown in Figure \ref{fig:UNet}. At first, the encoder is used to obtain and normalize the transformation of the input volume, using a Leaky ReLU activation function at each layer. At the bottleneck of this architecture, the volume will be in the size of 2×2×2 which represents the reduction of dimensionality prior to using a sigmoid activation function for segmentation. The decoder then up-samples this transformed 2×2×2 volume to reconstruct the image with this segmentation.
    
    Our main evaluation of the algorithm is through a combination of the Binary Cross Entropy Loss and Sørensen-Dice Loss given by the Dice Coefficient. For each segmentation method, we will tune the hyperparameters to minimize our loss function. 

\subsection{Sorensen-Dice Coefficient}
     The dice coefficient is the industry and academic standard for evaluating medical image segmentation and classification results. Evaluation is given through a scale of 0 to 1. Given two images $X$ and $Y$, a zero dice coefficient indicates that there is no similarity between two images while a one dice coefficient indicates that the images are exactly the same down to the pixel. The equation for the metric is given by 

    \begin{equation*}
        \text{DCS} = \frac{2|X \cap Y|}{|X| + |Y|}
    \end{equation*}
    
    From the literature review, most state-of-the-art dice scores on medical images range from 0.3 to 0.7. Much like a regression metric, a dice score of 0 or 1 are usually causes for concern since no two images from real-world data are truly the same. Although accuracy is often used as an evaluation metric, it is important to note that the dice score may be low even if the training and validation accuracy are high on a given dataset. 
    
\begin{figure}
        \centering
        \includegraphics[width = 70mm]{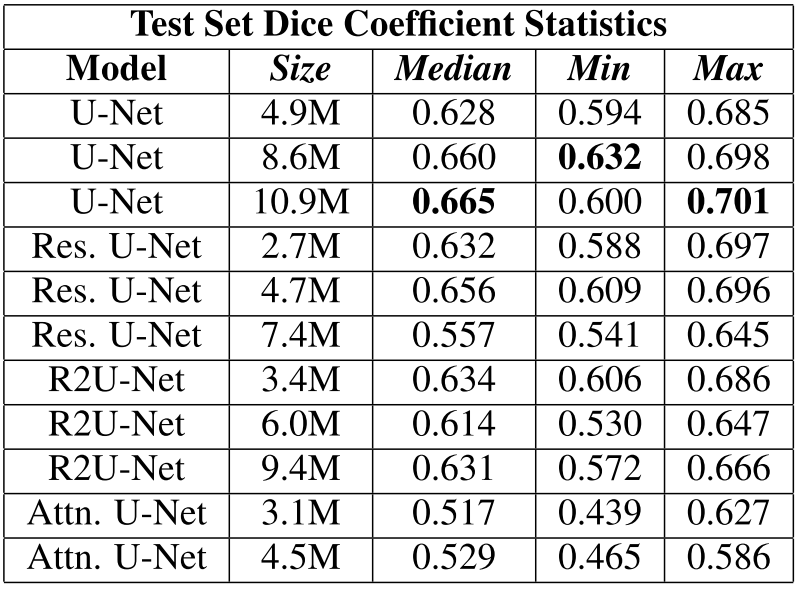}
        \caption{Adorno et.al. Dice Values \cite{Adorno2021AdvancingVision}}
        \label{fig:AdornoDice}
    \end{figure}

    Once again, our dataset is 514 images from the GI Data Science Lab and sized 512x512x3. To improve upon Adorno, et. al. \cite{Adorno2021AdvancingVision}, we augment the data through flipping and rotation for each image. This means that for each input image, we have created a flipped and rotated image as well, effectively tripling our input dataset. Data augmentation here makes the model generalize better due to the larger amount of training data. For reference, the results from Adorno, et. al. \cite{Adorno2021AdvancingVision} are shown in Figure \ref{fig:AdornoDice}. The size field in Figure \ref{fig:AdornoDice} refers to the total number of parameters of each model. For comparison, our UNet results are shown in Table \ref{tab:testDice}.

    \begin{table}[htbp]
      \centering
        \begin{tabular}{|c|c|c|c|c|}
        \hline
        \textbf{Model} & \textit{\textbf{Size}} & \textit{\textbf{Median}} & \textit{\textbf{Min}} & \textit{\textbf{Max}} \\
        \hline
        MCD UNet & 494K  & 0.591 & 0.470 & 0.650 \\
        \hline
        UNet & 494K & 0.598 & 0.442 & 0.657\\
        \hline
        DenseNet & 7.2M & 0.592 & 0.369 & 0.645\\
        \hline
        ResNet & 24M & 0.612 & 0.505 & 0.651\\
        \hline
        \end{tabular}%
      \caption{Test Results Dice Score}
      \label{tab:testDice}%
    \end{table}%
    
    Most significantly, we can see that our approach has only $\approx$ 494,000 parameters while the smallest model in Adorno et al. had 3.1 million parameters and the largest had 10.9 million parameters. Thus, our work is at least one order of magnitude less in size than Adorno et al. which means our model is less complex. Given that the same UNet approach, we can assume, holding all other factors such as GPU availability constant, our approach runs faster and more efficiently. Despite the significant difference is model size, our performance is similar to Adorno et. al with dice scores around 0.6, matching their Residual UNet and R2UNet results while outperforming their Attention UNet results.

\begin{figure}
        \centering
        \includegraphics[width = 90mm]{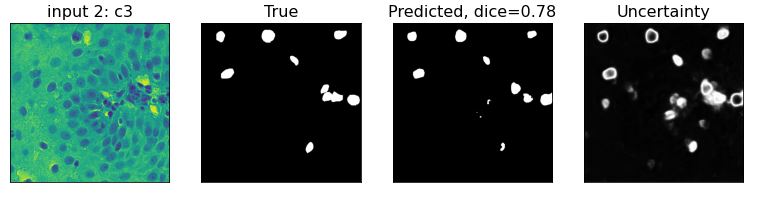}
        \caption{Output comparing the true, predicted, and uncertainty values for an image patch}
        \label{fig:uncertaintyimage}
    \end{figure}

    Possibly the strongest result of this approach is the visualization possible due to the quantification of our model's uncertainty shown in Figure~\ref{fig:uncertaintyimage}. 

    In this example image, white indicates high amounts of uncertainty while black indicates low amounts of uncertainty. We can see that the pixels are nearly white around the borders of the eosinophils essentially "highlighting" them in the output image. The high uncertainty on boundary pixels around each eosinophil indicate that until the model sees more of the eosinophil, the model hesitates to classify the pixels as an eosinophil. We can see this verified that once the model passes the boundary pixels of each eosinophil, the uncertainty drops significantly and the interior of the eosinophil is nearly black. This dramatic shift in values from the outside of the eosinophil which is black, to the border of the eosinophil which is white, to the interior of the eosinophil which is black again creates these "rings" of white in the resulting image. In areas where multiple eosinophils are clustered, the model seems to struggle to differentiate between the eosinophils leading to a "cloud" of high uncertainty around the area. However, we can see that this output provides valuable information to pathologists as the highlighting the general area of interest and reducing the visual load compared to the original input on the far left.

\section{Analysis and Interpretation}

\begin{figure}
        \centering
        \includegraphics[width = 70mm]{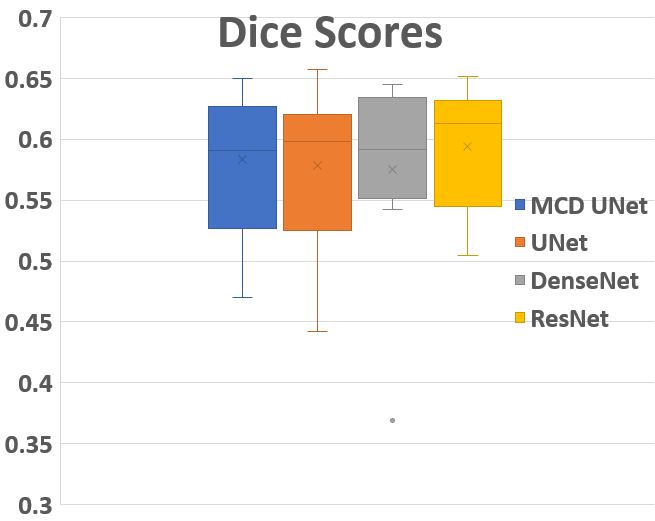}
        \caption{Boxplots of Model Dice Scores}
        \label{fig:dicebox}
    \end{figure}

For analysis, we plotted the Dice scores of all models in Figure~\ref{fig:dicebox}. The boxplots overlapped, which indicates the results from each neural net are not statistically significantly better than others. This is important because we can obtain uncertainty visualizations from the MCD UNet, as shown in Figure~\ref{fig:uncertaintyimage}, whereas UNet and ResNet50 do not allow for this. Monte Carlo Dropout allows us to visualize the uncertainty through the dropout layers since it randomly turns off neurons during training, which adds this stochastic element. 
    
As the novel piece in this work, we compared the model's uncertainty first through an overview of the raw data and through a boxplot for visualization. We can see that the uncertainty of almost all model is comparatively similar with only the DenseNet having an outlier uncertainty value of 0.05. Considering the significant difference in the order of magnitude between the sizes of the models, the relative similarity in the uncertainty values demonstrate that at least at a sufficient size, model uncertainty stays consistent regardless of the size of the model chosen.

    \begin{table}[htbp]
      \centering
        \begin{tabular}{|c|c|c|c|c|}
        \hline
        \textbf{Model} & \textit{\textbf{Size}} & \textit{\textbf{Median}} & \textit{\textbf{Min}} & \textit{\textbf{Max}} \\
        \hline
        MCD UNet & 494K  & 0.007 & 0.004 & 0.016 \\
        \hline
        UNet & 494K & 0.007 & 0.005 & 0.013\\
        \hline
        DenseNet & 7.2M & 0.009 & 0.006 & 0.05\\
        \hline
        ResNet & 24M & 0.008 & 0.005 & 0.01\\
        \hline
        \end{tabular}%
      \caption{Model uncertainty}
      \label{tab:testuncertainty}%
    \end{table}%

    \begin{figure}
        \centering
        \includegraphics[width = 70mm]{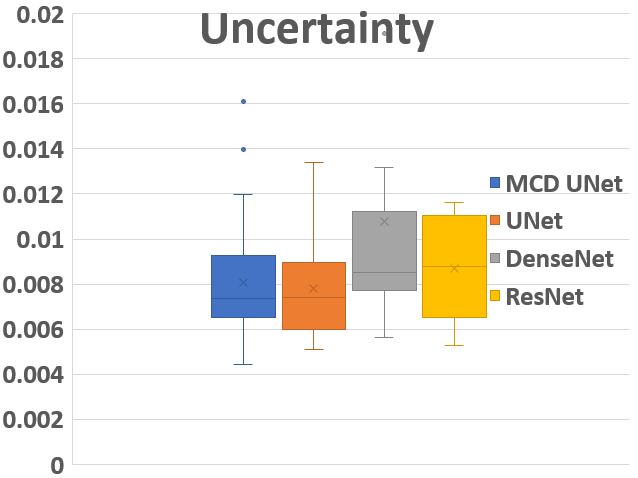}
        \caption{Boxplots of Model Uncertainty}
        \label{fig:uncertaintybox}
    \end{figure}

\section{Conclusion and Future Directions}

As stewards of patient data, researchers in medical image analysis must ensure all models perform efficiently and accurately. This work addresses both concerns demonstrating MCD UNet's comparable performance at an order of magnitude smaller than current models and introducing model uncertainty as an evaluation metric. All models in this work had comparable values in uncertainty indicating that at least after a model reaches a certain size, the aleatoric uncertainty will stay constant. We then illustrated the model's uncertainty through a visualization which highlighted the eosinophils in the resulting image. Scaling this approach with uncertainty to full size biopsy images can help pathologists quickly identify eosinophils while also reducing the mental load. Compared to a screen full of cells and color, the black and white "rings" circling the eosinophils can at least narrow down eosinophil locations while having the potential to count all eosinophils and output a mask showing their exact locations. One of the most difficult parts of this work was working with limited patient data which is a common challenge in the field of medical image analysis. A potential improvement to this work would be to incorporate few-shot learning. While our dataset exists in a high-dimensional space, we only have a limited number of samples available to us due to the cumbersome nature of acquiring annotated histology images. Few-shot learning has significantly improved classification accuracy in medical imaging datasets \cite{Kim2017Few-shotDiagnosis} and likely would produce competitive results.

\begin{acks}
The authors would like to thank the medical researchers of the UVA Gastroenterology (GI) Data Science Laboratory for obtaining the EoE dataset through biopsies of UVA Medical Center patients. Additionally, the staff of the GI Data Science Laboratory tirelessly provided direct medical feedback on our results. Research reported in this publication was supported by National Institutes of Health (NIH) through the National Institute of Diabetes and Digestive and Kidney Diseases (NIDDK) under award numbers K23DK117061-01A1 (Syed) and R01DK132369 (Syed).
\end{acks}

\bibliographystyle{ACM-Reference-Format}
\bibliography{references}

\appendix

\end{document}